\PassOptionsToPackage{bookmarks=false}{hyperref}
\documentclass[sigconf]{acmart}

\usepackage{booktabs} 
\usepackage{fancyvrb}
\usepackage{fancyvrb}
\usepackage{multirow}
\usepackage{subfigure}
\usepackage{amsmath}
\usepackage{amssymb}
\usepackage[linesnumbered,ruled,vlined]{algorithm2e}
\usepackage{hyperref}
\usepackage{hyphenat}
\usepackage{enumitem}
\newcommand{\texttildemid}{\raisebox{0.5ex}{\textasciitilde}}

\renewcommand\footnotetextcopyrightpermission[1]{} 
\begin{document}

\settopmatter{printacmref=false}
\renewcommand\footnotetextcopyrightpermission[1]{} 
\title{Automated Localization for Unreproducible Builds}
\author{Zhilei Ren} 
\affiliation{Key Laboratory for Ubiquitous Network and Service Software of Liaoning Province,
School of Software, \\
Dalian University of Technology, Dalian, China}
\email{zren@dlut.edu.cn}
\author{He Jiang}
\affiliation{Key Laboratory for Ubiquitous Network and Service Software of Liaoning Province,
School of Software, \\
Dalian University of Technology, Dalian, China}
\email{jianghe@dlut.edu.cn}
\author{Jifeng Xuan}
\affiliation{School of Computer Science,\\ Wuhan University,\\ Wuhan, China}
\email{jxuan@whu.edu.cn}
\author{Zijiang Yang}
\affiliation{Department of Computer Science,\\ Western Michigan University,\\ Kalamazoo, MI, USA}
\email{zijiang.yang@wmich.edu}

\begin{abstract}
Reproducibility is the ability of recreating identical binaries under pre-defined build environments.  Due to the need of quality assurance and the benefit of better detecting attacks against build environments, the practice of reproducible builds has gained popularity in many open-source software repositories such as Debian and Bitcoin. However, identifying the unreproducible issues remains a labour intensive and time consuming challenge, because of the lacking of information to guide the search and the diversity of the causes that may lead to the unreproducible binaries. 

In this paper we propose an automated framework called RepLoc to localize the problematic files for unreproducible builds. RepLoc features a query augmentation component that utilizes the information extracted from the build logs, and a heuristic rule-based filtering component that narrows the search scope. By integrating the two components with a weighted file ranking module, RepLoc is able to automatically produce a ranked list of files that are helpful in locating the problematic files for the unreproducible builds. We have implemented a prototype and conducted extensive experiments over 671 real-world unreproducible Debian packages in four different categories. By considering the topmost ranked file only, RepLoc achieves an accuracy rate of $47.09\%$. If we expand our examination to the top ten ranked files in the list produced by RepLoc, the accuracy rate becomes $79.28\%$. Considering that there are hundreds of source code, scripts, Makefiles, etc., in a package, RepLoc significantly reduces the scope of localizing problematic files. Moreover, with the help of RepLoc, we successfully identified and fixed six new unreproducible packages from Debian and Guix.

\end{abstract}


\maketitle

\section{Introduction}
\label{intro}
As an indicator of the ability that the binaries could be recreated consistently from source, recent years have witnessed the emerging idea of reproducible builds.  
Given the source files, the reproducibility is described as the ability of building identical binary under pre-defined build environments \cite{debianabout}. 
In this study, source files include source code, scripts, Makefiles, build configurations, etc \cite{debianpackage}.  
Checking the reproducibility of software creates a verifiable linkage that bridges the gap between the readable source files and the binary packages, which is important from various perspectives.

Firstly, reproducibility is very important for the safety of build environments. For software ecosystems, attacks against the build environment may lead to serious consequences. By compromising the system to produce packages with backdoors \cite{wheeler2005countering, holler2015evaluation}, malicious behaviors such as trusting trust attack \cite{thompson1984reflections} may be introduced during the build time. For example, in 2015, over 4,000 iOS applications were infected by a counterfeit version of Apple's Xcode development environment (known as XcodeGhost) \cite{fireeye}. XcodeGhost injected malicious code during compiling time so that developers unknowingly distributed malware embedded in their applications \cite{xcodeghost}. Obviously, a solution is to ensure that the same source files always lead to the same binary packages so that an infected different binary immediately raises alarms. Unfortunately, a major obstacle of detecting such attacks lies in the transparency gap between the source files and their compiled binary packages. Due to non-deterministic issues such as timestamps and locales, it is not uncommon that rebuilding an application yields different binaries even within secure build environments. Therefore, these kinds of attacks often elude detection because different binaries of the same application is normal. 

\begin{figure}[!t]
	\includegraphics[width=.45 \textwidth]{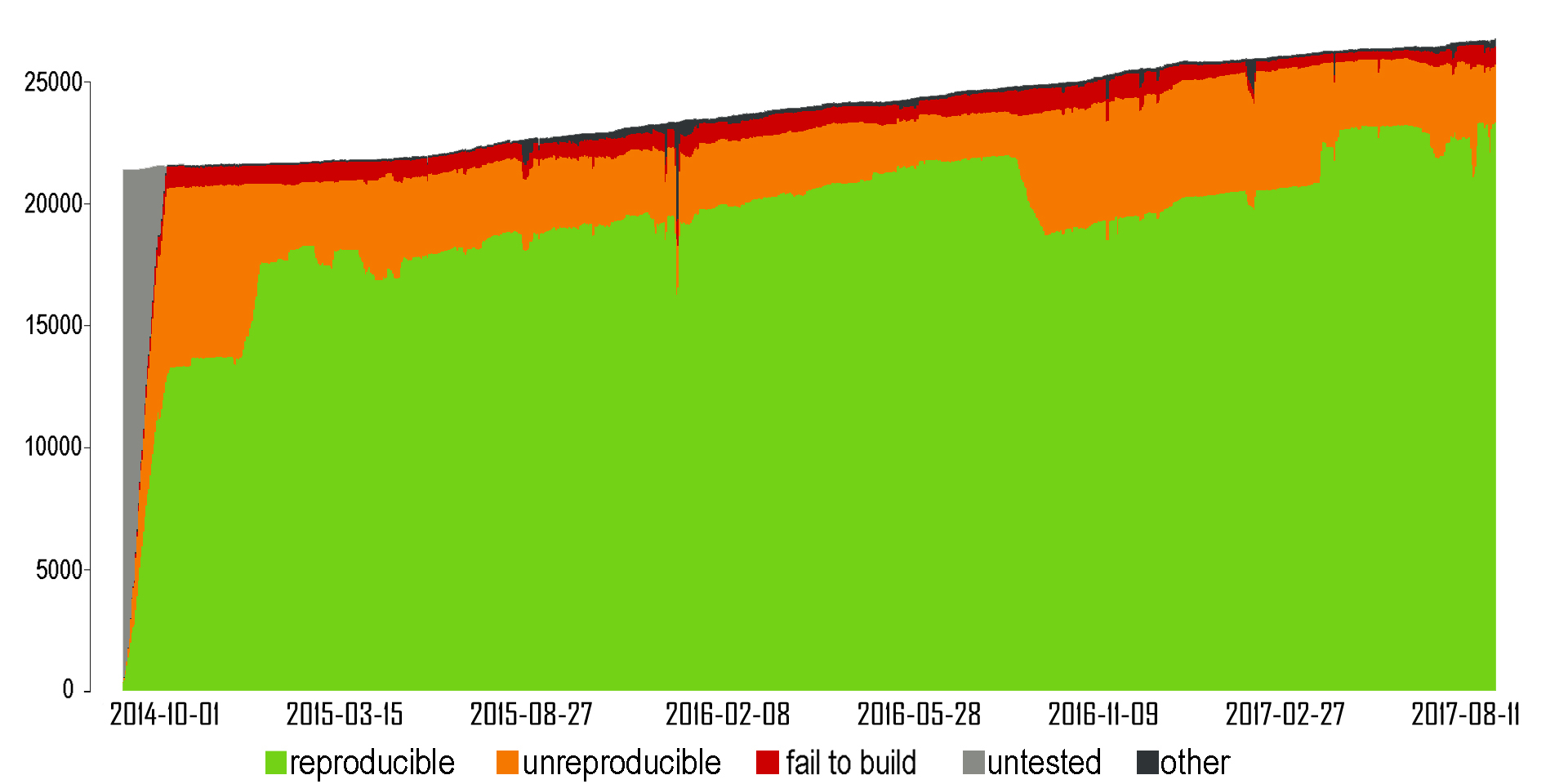}
	\caption{Reproducibility status of Debian unstable for AMD64}
	\label{repro-status}
\end{figure}

Besides detecting attacks against build environments, validating the reproducibility is also helpful in debugging and finding certain release-critical bugs (e.g., \texttt{libical-dev} 1.0-1.1) \cite{bugreport}. Furthermore, in the context of and continuous integration and software upgrade \cite{DIRUSCIO2014438}, reproducible packages could be helpful in caching, and reducing redundant operations, e.g., by eliminating the necessity of delivering the different binaries compiled from the same source files. Due to the significant benefits, many open-source software repositories have initiated their validation processes. These repositories include GNU/Linux distributions such as Debian and Guix, as well as software systems like Bitcoin \cite{who}. For instance, since 2014, the number of Debian's reproducible packages has been steadily increasing. Figure~\ref{repro-status} presents the trend of the reproducible builds in Debian \cite{debianstatus}. As of August 2017, over 85\% of Debian's packages could be reproducibly built.  

Despite the effort towards reproducibility, many packages remain unreproducible. For example, according to Debian's Bug Tracking System (BTS), as of August 23, 2017, there are 2,342 packages that are not reproducible \cite{debianstatus} for the unstable branch targeting the AMD64 architecture. Such large number of unreproducible packages implies the challenges in detecting and then fixing the unreproducible issues. In particular, the localization task for the problematic files is the activity of identifying the source files that cause unreproducibility, which ranks source files based on their likelihood of containing unreproducible issues. Currently, the localization task is mostly manually conducted by developers. Since there may be hundreds to thousands of source files for a package, the localization tends to be labor intensive and time consuming. 

To address this problem, we consider the source files as text corpus, and leverages the diff log\footnote{Generated by \texttt{diffoscope}, \url{https://diffoscope.org}} generated by comparing the different binaries to guide the search. As such, the localization of the problematic files can be modeled as a classic Information Retrieval (IR) problem: given the source files and the diff log, determine those problematic files from the source files that are relevant to the unreproducible issues. The IR model has the potential to automate the localization task. However, the localization task is challenging, due to its unique characteristics.

First, the information for locating the problematic files within the source files is very limited. The diff log generated by comparing the different binaries, which is considered as the input of the IR process, may not be sufficiently informative.  We call this challenge an \textbf{information barrier}. In addition, there are many causes that may lead to unreproducible builds, such as embedding timestamps in files and recording file lists in non-deterministic order. The detailed issues are manually listed in Debian's documentation \cite{knownissues}. Moreover, the diverse types of files in a package also add to the complexity of localizing the problematic files, which may reside in not only the source code, but also other types of files such as scripts, Makefiles and build configurations. We call this challenge a \textbf{diverse-cause barrier}.

To break through the barriers, we propose a localization framework called RepLoc that targets the localization task in search of problematic files for unreproducible builds. Given an unreproducible package with two different built binaries as the input, RepLoc produces a list of ranked source files. RepLoc features two components that address the two aforementioned challenges. For the information barrier, we develop a Query Augmentation (QA) component that utilizes the information extracted from the build logs to enhance the quality of the queries (represented by the file names extracted from the diff logs, see Section \ref{background}). 
For the diverse-cause barrier, we develop a Heuristic rule-based Filtering (HF) component. More specifically, we propose 14 heuristic rules that are obtained by summarizing the information presented in Debian's documents. Furthermore, we employ a weighted File Ranking (FR) component to combine the QA and HF components, and build an integrated framework to automate the localization of the problematic files for unreproducible builds.

To evaluate RepLoc, we have collected a real-world dataset that consists of 671 unreproducible packages. Since these packages were later fixed with patches from Debian's BTS, we know exactly which files caused the unreproducibility and thus can use the facts to evaluate the accuracy of RepLoc. If we consider the topmost ranked file only, RepLoc achieves an accuracy rate of $47.09\%$. If we expand the range to include top ten ranked files, the accuracy rate becomes $79.28\%$. For other metrics such as precision and recall, RepLoc also outperforms the comparative approaches significantly. To further evaluate the effectiveness of our approach, we use RepLoc on unreproducible packages that have never been fixed before. With the help of RepLoc, we successfully identified the problematic files, then manually fixed the unreproducible issues over three Debian packages. Moreover, the usefulness of RepLoc is examined over a different software repository (Guix \cite{guix} in this study). Under the guidance of RepLoc, problematic files for three unreproducible packages from Guix are detected and fixed.

This paper makes the following main contributions.
\begin{itemize}[leftmargin=*]
		\item To the best of our knowledge, this is the first work to address the localization task for unreproducible builds.
		\item We propose an effective framework RepLoc that integrates heuristic filtering and query augmentation. A prototype has been implemented based on the approach. 
		\item We have evaluated RepLoc on 671 unproducibile packages that were later fixed in the Debian repository. The experimental results show that RepLoc is effective. We have made the benchmarks publicly available at \textbf{\url{https://reploc.bitbucket.io}}.
		\item  Under the guidance of RepLoc, we fixed six unreproducible packages from Debian and Guix, and submitted the patches to the BTSs of the two repositories. Among the submitted patches, four have been accepted.
\end{itemize}

The rest of this paper is organized as follows. In Section~\ref{background}, we give the background of this work. Our approach is presented in Section \ref{framework}, followed by experimental study in Section \ref{experiment}. The threats to validity and related work are described in Sections \ref{threats}--\ref{related}. Finally, Section \ref{conclusion} concludes the paper. 

\section{Background}
\label{background}

\begin{figure}[!t]
	\centering
	\includegraphics[width=.35 \textwidth]{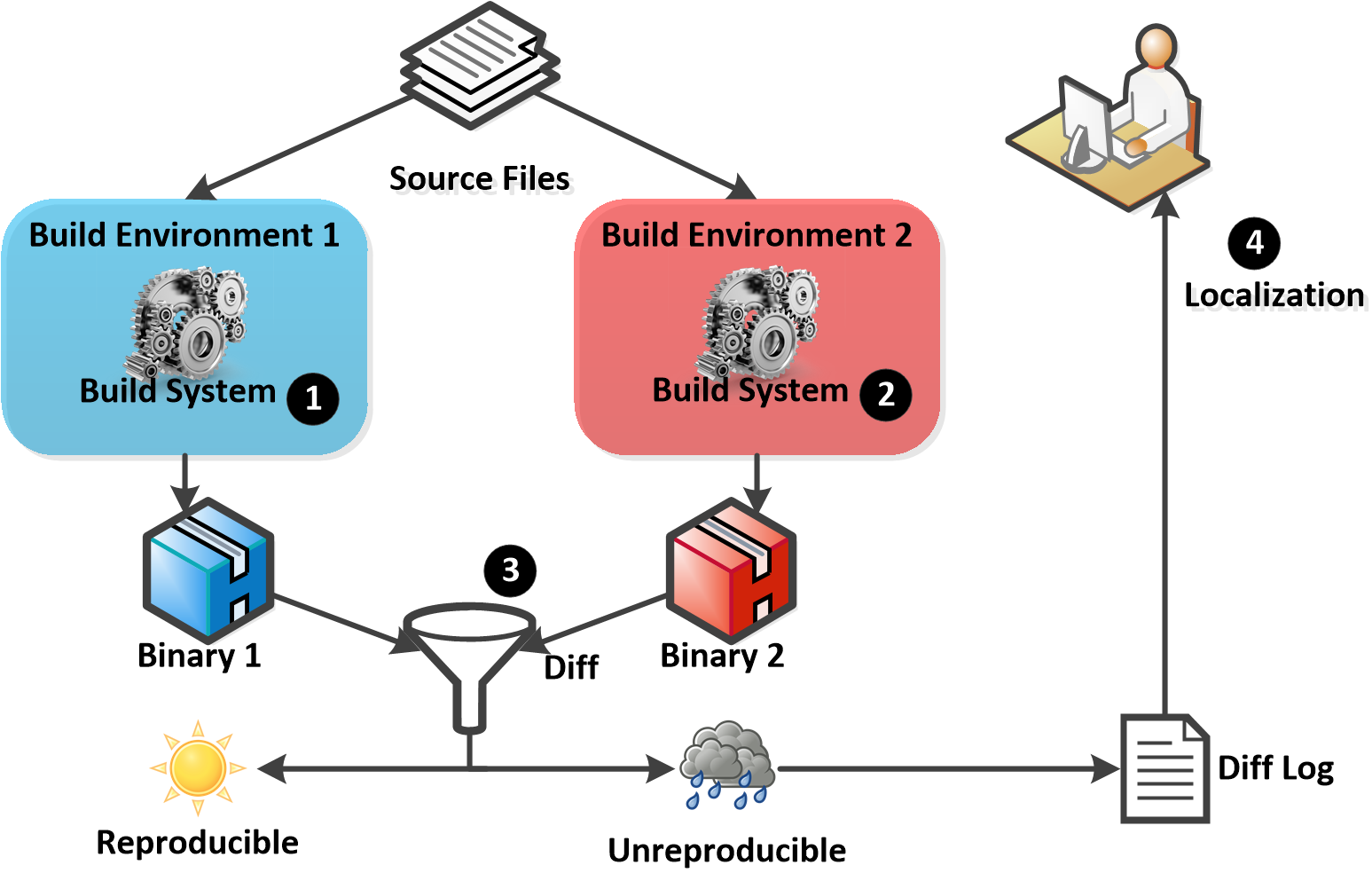}
	\caption{Reproducibility validation work flow}
	\label{flow}
\end{figure}
\begin{table}[t]
	\centering
	\caption{Snippet of altered environment variations}
	\label{environment}
	\scriptsize
	\begin{tabular}{lll}
		\hline
		Configuration	&First build	&Second build\\
		\hline
env TZ                      & ``/usr/share/zoneinfo/Etc/GMT+12'' & ``/usr/share/zoneinfo/Etc/GMT-14''\\
env LANG                    & ``C''                              & ``fr\_CH.UTF-8''\\
env LANGUAGE                & ``en\_US:en''                      & ``fr\_CH:fr''\\
env BUILDDIR                & ``/build/1st''                     & ``/build/2nd''\\
\ldots                      & \ldots                             & \dots\\
\hline
	\end{tabular}
\end{table}

Taking Debian as a typical example, Figure~\ref{flow} illustrates the common work flow of validating the reproducibility of packages \cite{experimental}. First, the source files are compiled under two pre-defined build environments (steps 1--2). More specifically, the build environments are constructed by setting up altered environment variables or software configurations. For instance, within Debian's continuous integration system,\footnote{\url{https://jenkins.debian.net}} altered environment variables include locales, timezones, user privileges, etc. Table~\ref{environment} presents a snippet of the altered environment (see \cite{variant} for more detailed information). Two versions of binaries can be generated with respect to each environment. The two versions are then compared against each other (step 3). If they are not bit-to-bit identical, the localization of problematic files that lead to unreproducible builds is required, based on the diff log and the source files (step 4). 

The build and the comparison procedures (steps 1--3) can easily be automated, but the localization (step 4) mainly relies on the developers. Unfortunately, manual effort to identify the files that lead to unreproducible builds is nontrivial. As shown in Figure~\ref{flow}, the diff logs are the major source of the information to guide the localization of the problematic files, which, unfortunately, are not always sufficiently informative. 

\begin{figure}[t]
	\centering
	\includegraphics[width=0.5\textwidth]{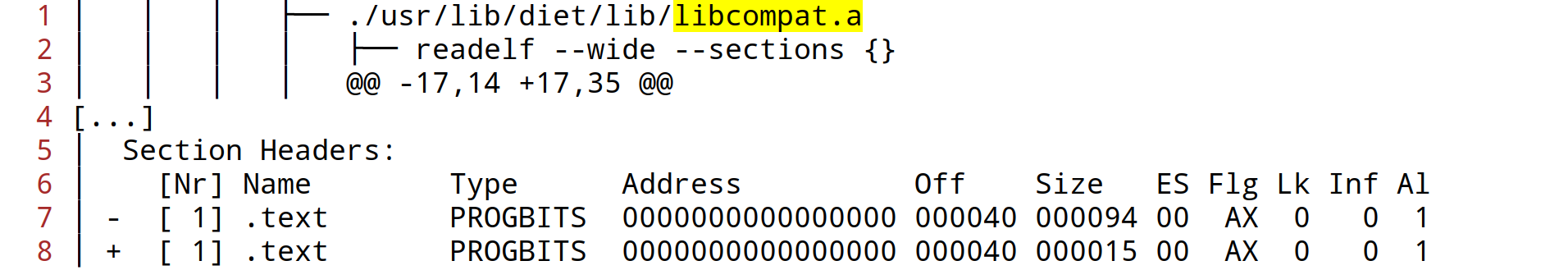}
	\caption{Diff log snippet for \texttt{dietlibc}}
	\label{diff-snippet}
\end{figure}

\begin{figure}[t]
	\centering
	\subfigure[Makefile snippet]{
	\includegraphics[width=0.5\textwidth]{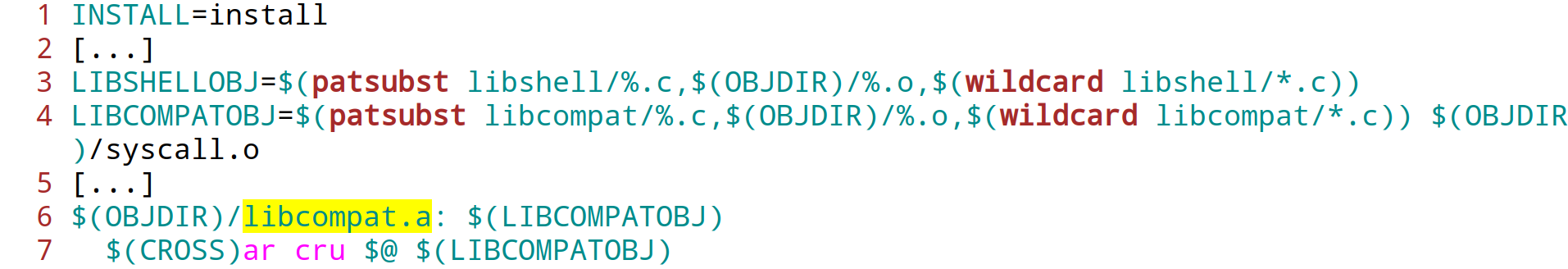}
	\label{make-snippet}
	}
	\subfigure[Patch snippet]{
	\includegraphics[width=0.5\textwidth]{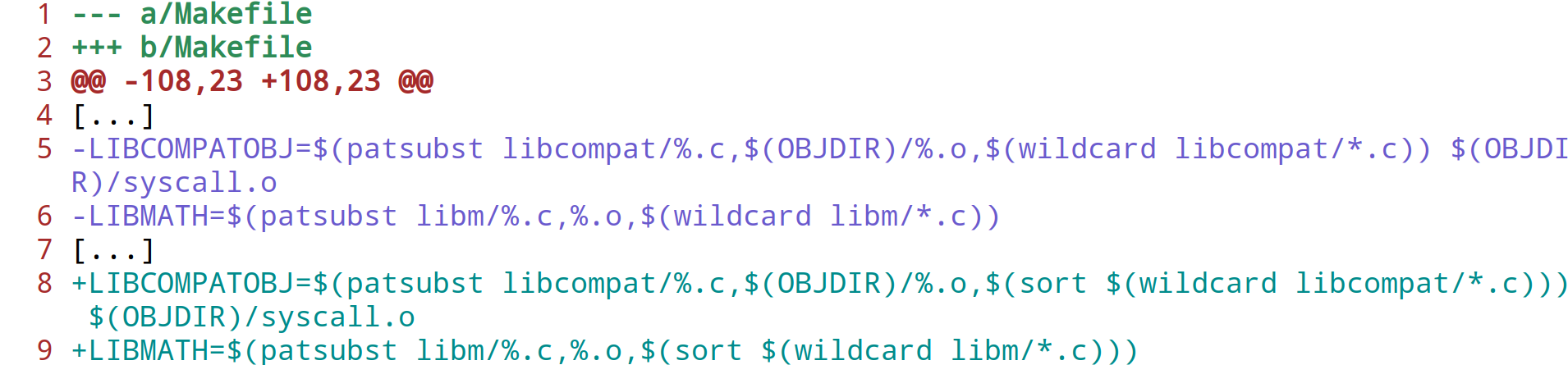}
	\label{patch-snippet}
	}
	\caption{Makefile and patch snippet for \texttt{dietlibc}}
	\label{make-patch}
\end{figure}

Figure~\ref{diff-snippet} gives a snippet of the diff log for \texttt{dietlibc}, a libc implementation optimized for small size. In the original version (0.33{\texttildemid}cvs20120325-6), a static library file  differs between the two versions during the build time (\texttt{/usr/lib/diet/lib/libcompat.a}). As shown in the diff log, \texttt{diffoscope} indicates the difference via the output of the GNU binary utility \texttt{readelf}. However, since the diff content may not be well comprehensible (e.g., lines 7--8 in Figure~\ref{diff-snippet}), we do not leverage such information in this study. Meanwhile, Figure~\ref{make-patch} presents a snippet of a problematic file (\texttt{/Makefile}) and the patch that fixes the issue. In Figure~\ref{patch-snippet}, line 8 indicates that the root cause of the unreproducibility lies in the non-stable order of the object files, which are fed to the \texttt{ar} utility to generate \texttt{libcompat.a} (lines 6--7 of Figure~\ref{make-snippet}). The difficulty in this example is that, the diff log may fail to provide sufficient information.  Though it is possible to match the correct file with only the file name, i.e., line 6 of Figure~\ref{make-snippet}, chances are that other irrelevant files containing the same file name might be matched as well.

The aforementioned example illustrates how problematic files can be detected and fixed. In reality there are multiple altered build configurations and can be many corresponding causes that lead to unreproducible builds. For example, changing the timezone environment variable (\texttt{env TZ}) may cause the C/C++ packages that embed \texttt{\_\_DATE\_\_} macro to be unreproducible, and the locale environment variable (\texttt{env LC\_*}) may trigger unreproducible issues of packages that capture the text generated by programs. These diverse unreproducible causes make the localization task difficult.

\section{Our Approach}
\label{framework}
\begin{figure}[t]
	\centering
	\includegraphics[width=.44 \textwidth]{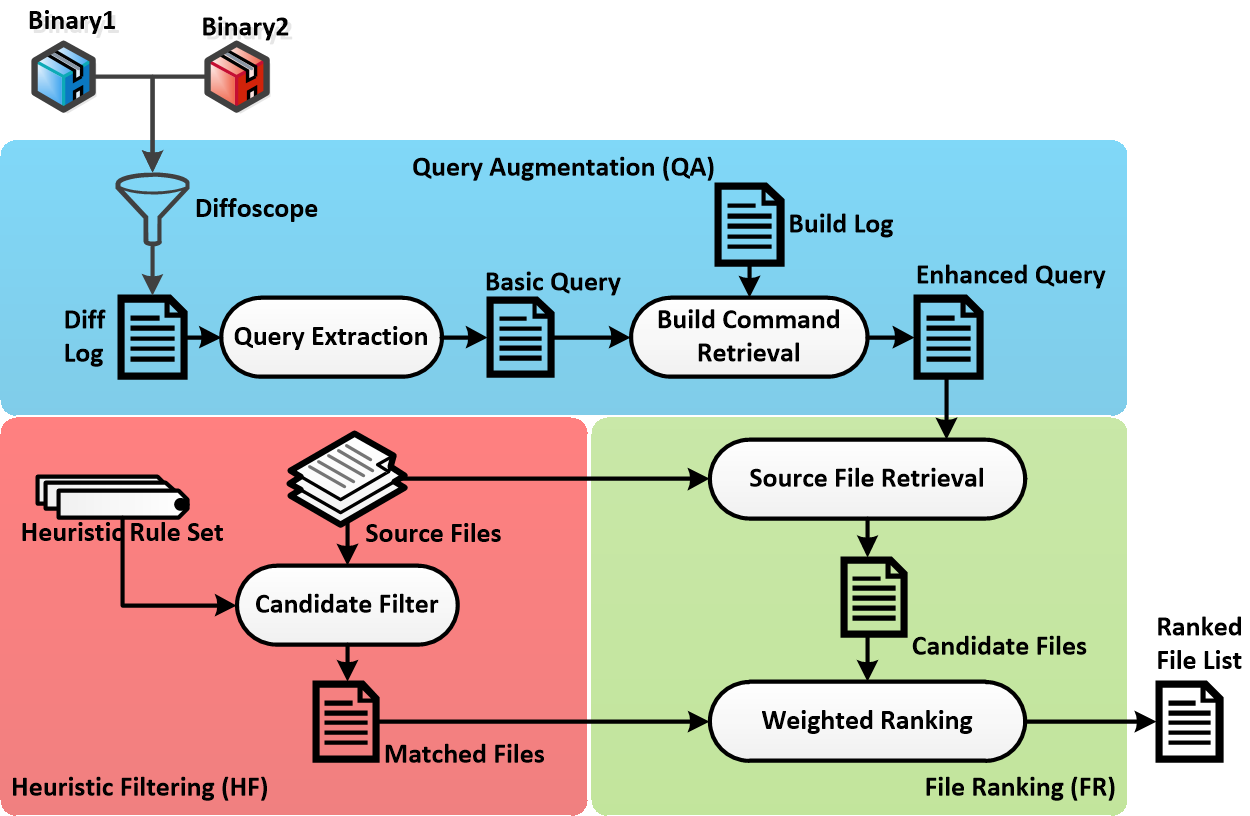}
	\caption{The RepLoc Framework}
	\label{repflow}
\end{figure}

In this section, we discuss the details of RepLoc. Figure~\ref{repflow} depicts the work flow of RepLoc that consists of three components QA, HF, and FR. For each component, we shall explain its design and implementation, companioned with the intermediate results over the running example \texttt{dietlibc}. 

\subsection{Query Augmentation Component}
\label{qa}

The upper part of Figure~\ref{repflow} depicts the QA component, which enriches the queried information by matching the files in the diff log and the build logs, to tackle the information barrier. 

First, the diff log is generated using \texttt{diffoscope}. Then, the query extraction module takes the diff log as the input, and generates the basic query. In this study, the basic query consists of the file names in the diff log. As mentioned, due to the information barrier, the information that can be utilized to localize the problematic files is limited other than a list of files that are different within the two build processes. Thus, we enhance the quality of the queries with the build command retrieval module. The motivation for this module is that, during the build process, the build information such as the executed commands can be obtained. Moreover, based on the co-occurrence relationship between the file names in the diff log and the build commands, we can identify the build commands with which the files mentioned in the diff log are built. Hence, it is rational to augment the query by supplementing the build commands from the build log.

\begin{figure}[!t]
	\centering
	\includegraphics[width=0.48\textwidth]{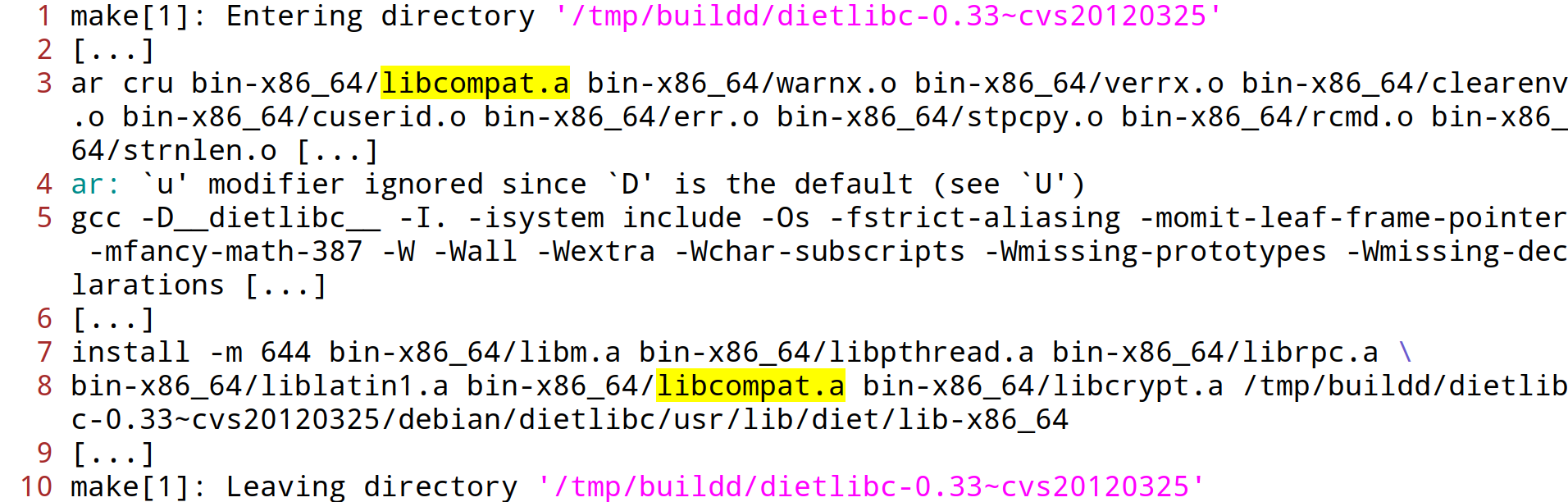}
    \caption{Build log snippet for \texttt{dietlibc}}
	\label{build-snippet}
\end{figure}
Figure~\ref{build-snippet} illustrates a snippet of the build log of the exemplifying package \texttt{dietlibc}. It can be observed that the build log is more informative and provides supplementary information with respect to the diff log. More specifically, we first split the build log into build command segments, with respect to the ``\texttt{Entering/Leaving directory}'' tags generated by \texttt{make} (e.g., lines 1 and 10 of Figure~\ref{build-snippet}). With this operation, the commands invoked under the same directory can be grouped together, as a file of the augmentation corpus (denoted as a command file). Note that though there are two versions of build logs with respect to the two build environments, since we are interested in the build command, the choice of either version of build log does not have an impact on the results. Then, the relevant files in the corpus are obtained by utilizing an IR model. In essence, any IR model can be adopted. In this study,  we employ the Vector Space Model (VSM), due to its simplicity and effectiveness. 

To realize the VSM based augmentation, we calculate the cosine similarity between basic query and the command files. Thereafter, the matched commands from the most relevant command files are obtained. In particular, for the VSM model, we assign weight value for each file with the \textit{TF-IDF} (Term  Frequency-Inverse  Document  Frequency) measurement, which is widely used in IR \cite{manning2008introduction}. In this paper, for a term $t$ in a document $d$, its $\textit{TF-IDF}_{t,d}$ value is calculated based on $f_{t,d} \times \frac{N}{n_t}$,
where $f_{t,d}$ indicates the number of $t$'s occurrences in $d$, $n_t$ denotes the number of files in which $t$ appears, and $N$ means the number of source files. With \textit{TF-IDF} defined, each file is represented as a vector, and the cosine similarity with the basic query is used to rank the command files.
\begin{equation}
	Sim(\vec l, \vec s) = \frac{\vec{l} \cdot \vec{s}}{|\vec{l}| | \vec{s}|},
	\label{sim1}
\end{equation}
where $\vec l \cdot \vec s$ represents the inner product of the basic query and the command file, and $|\vec{l}| | \vec{s}|$ denotes the product of 2-norm of the vectors.
After that, the basic query and the retrieved contents, which are commands executed during the build process, are concatenated together as the enhanced query.

\textbf{Running example:} For \texttt{dietlibc}, all the file names in the diff log, e.g., \texttt{./usr/lib/diet/lib/libcompat.a}, are extracted as the basic query. Then, within the augmentation, \texttt{ar cru bin-x86\_64/libcompat.a [\dots]} (line 3 of Figure~\ref{build-snippet}) and the build commands in the same command file are retrieved. Finally, the contents of the retrieved command files are appended after the basic query, as the final query.

\begin{table}[t]
	\caption{Heuristic rule set}
	\label{rules}
	\centering
	\footnotesize
	\begin{tabular}{|l|l|l|} 
		\hline
		ID& Rule & PCRE statement \\
		\hline
1  & \texttt{TIME\_MACRO}         & \textbf{\_\_TIME\_\_} \\
2  & \texttt{DATE\_MACRO}         & \textbf{\_\_DATE\_\_} \\
3  & \texttt{GZIP\_ARG}           & \verb|\|b\textbf{gzip}\verb|\|s(?!.*-[a-z9]*n) \\
4  & \texttt{DATE\_CMD}           & (\verb|\|\$\verb|\|(\textbf{date})$|$(\verb|\|\$\verb|\|(shell\verb|\|s*\textbf{date})$|$(\verb|\|\`{}\textbf{date}) \\
5  & \texttt{PY\_DATE}            & \textbf{datetime}\verb|\|.\textbf{datetime}\verb|\|.\textbf{today} \\
6  & \texttt{PL\_LOCALTIME}       & \verb|\|\$\verb|\|.*\textbf{localtime} \\
7  & \texttt{SYSTEM\_DATE}        & \textbf{system.*date} \\
8  & \texttt{DATE\_IN\_TEX}       & \verb|\|\verb|\|\textbf{date}.*\verb|\|\verb|\|\textbf{today} \\
9 & \texttt{SORT\_IN\_PIPE}      & \^{}.*\verb|\|$|$`(?!.*LC\_ALL=).*\verb|\|s*\textbf{sort}\verb|\|b \\
10 & \texttt{GMTIME}              & \textbf{gmtime}\verb|\|( \\
11 & \texttt{TAR\_GZIP\_PIPE}     & \verb|\|b\textbf{tar}\verb|\|b.*\verb|\|$|$\verb|\|s*\verb|\|b\textbf{gzip}\verb|\|b \\
12 & \texttt{PL\_UNSORTED\_KEY}   & (\^{}(?!.*\textbf{sort}).*\verb|\|s*\textbf{keys}\verb|\|s*\%) \\
13 & \texttt{LS\_WITHOUT\_LOCALE} & \^{}.*\verb|\|\$\verb|\|(.*(?!.*LC\_ALL=).*\verb|\|s*\verb|\|b\textbf{ls}\verb|\|b \\
14 & \texttt{UNSORTED\_WILDCARD}  & (\^{}(?!.*\textbf{sort}).*\verb|\|s*\verb|\|b\textbf{wildcard}\verb|\|b) \\ 
		\hline
	\end{tabular}
\end{table}

\subsection{Heuristic Filtering Component}
\label{hf}

The HF component is designed to capture the problematic files by incorporating the domain knowledge, which is represented as frequently observed patterns. In HF, the heuristic rules are constructed based on the following criteria: (1) The rules are manually constructed based on Debian's documentation \cite{notes}. (2) The rules are summarized for the four major categories of unreproducible issues (see Setcion \ref{preparation}). We traverse the notes in the documentation, and capture those issues that are described as Perl Compatible Regular Expression (PCRE). For example, invoking \texttt{gzip} without ``-n'' argument could be expressed using the negative assertions feature of PCRE (rule 3 in Table \ref{rules}). Meanwhile, as a counterexample, the timestamps embedded in  Portable Executable (PE) binaries are hard to be identified by heuristic rules or even by developers \cite{timestamppe}.  
After manual inspection based on the criteria, we obtain 14 heuristic rules, which are presented in Table~\ref{rules}, and described as follows:

(1) \texttt{TIME\_MACRO}: using C time preprocessing macro in source files will embed different timestamps when compiled at different times.  (2) \texttt{DATE\_MACRO}:~embedding C date preprocessing macro in source files is similar as the previous case.  (3) \texttt{GZIP\_ARG}: if applying gzip without -n argument, timestamps will be embedded in the header of the final compressed file.  (4) \texttt{DATE\_CMD}: capturing the current date with the \textbf{date} shell command.  (5) \texttt{PY\_DATE}: obtaining date time in Python scripts.  (6) \texttt{PL\_LOCALTIME}: obtaining date time in Perl scripts.  (7) \texttt{SYSTEM\_DATE}: recording system time in the compiled binary.  
(8) \texttt{DATE\_IN\_TEX}: embedding date in TeX files, which influences the built pdf files.  (9)  \texttt{SORT\_IN\_PIPE}: execute \textbf{sort} in pipeline without locale setting.  (10)  \texttt{GMTIME}: obtaining current date time.  (11)  \texttt{TAR\_GZIP\_PIPE}: execute \textbf{tar} and \textbf{gzip} in pipeline.  (12)  \texttt{PL\_UNSORTED\_KEY}: traversing unsorted hash keys in Perl script does not guarantee identical order.  (13)  \texttt{LS\_WITHOUT\_LOCALE}~: capturing \textbf{ls} without locale setting is similar with \texttt{SORT\_IN\_PIPE}.  (14)  \texttt{UNSORTED\_WILDCARD}: using \textbf{wildcard} in Makefiles without sorting, similar with \texttt{PL\_UNSORTED\_KEY}.

By applying the rules over the source files (e.g., with GNU \texttt{grep} \texttt{-r -P}), we obtain a subset of files that may lead to unreproducible builds. Note that these rules equally treat the source files as plain text, rather than consider the file types (e.g., based on file extension). The reason is that the unreproducible issues may reside in snippets or templates that do not follow file extension conventions, which are eventually embedded into unreproducible binaries. Based on such consideration, we do not sort the matched files in HF.

\textbf{Running example:} For \texttt{dietlibc}, there are in total five problematic files, namely, \texttt{/libpthread/Makefile}, \texttt{/libdl/Makefile}, \texttt{/debian/\{rules}, \texttt{implicit\}}, and \texttt{/Makefile}. Among these files, \texttt{/Makefile} (see Figure~\ref{patch-snippet}) can be captured by the \texttt{UNSORTED\_WILDCARD} rule, in which \texttt{sort} does not appear before \texttt{wildcard}. However, we should note that there may be false alarms, e.g., for unexecuted commands or text in the comments. Consequently, HF may fail to place the matched problematic files at the top of the list.

\begin{algorithm}[t]
\footnotesize
\DontPrintSemicolon
\SetAlgoLined
\KwIn{binary package \textit{first}, binary package \textit{second}, weight $\alpha$}
\KwOut{candidate file list \textit{result}}
\Begin
{
    \tcp{Query Augmentation}
	$\textit{log} \leftarrow \texttt{diffoscope} (\textit{first}, \textit{second})$\;
	$\textit{query} \leftarrow \textit{parse\_log}(\textit{log})$\;
	$\textit{command\_files} \leftarrow \textit{parse\_build\_log}(\textit{build\_log})$\;
	$\textit{relevant\_command} \leftarrow \textit{retrieve\_relevant}(\textit{query}, \textit{command\_files})$\;
	$\textit{augmented} \leftarrow \textit{concatenate}(\textit{query}, \textit{relevant\_commant})$\;

    \tcp{Heuristic Filtering}
	$list \leftarrow \emptyset$\;
	\For {\textrm{\upshape each source file $s$}}
	{
		\lIf {\textrm{\upshape $s$ is matched by any rule in Table~\ref{rules}}}
		{
			$\textit{list} \leftarrow \textit{list } \cup \{s\}$ 
		}
	}
    \tcp{File Ranking}

	\For {\textrm{\upshape each source file $s$}}
	{
		\lIf {$s \in \textit{list}$}
		{ $w_s \leftarrow 1$ } 
		\lElse 
		{ $w_s \leftarrow 0$ }
		$\textit{score\textsubscript{s}} \leftarrow$ Calculate $Sim^\prime$ with respect to Equation~\ref{sim2}\;
	}
	\Return $\textit{sort}(\textit{source\_files}, \textit{score})$\;
}
\caption{RepLoc}
\label{algo-reploc}
\end{algorithm}

\subsection{File Ranking Component}
\label{ir}
The motivations behind the combination of HF and QA are twofold: (1) The heuristic rules in HF focus on the static aspect of the source files, i.e., treat all the source files in a unified way, and capture the suspicious files that match the defined patterns. Such mechanism can handle various file types. Unfortunately, there may be false alarms, especially for those files unused during the build process. (2) The build log based augmentation takes the dynamic aspect of the build process into consideration. With QA, we concentrate on the commands invoked during the build process. Hence, by combining the mechanisms, we can strengthen the visibility of the problematic files that lead to unreproducible builds. 

In the FR component, these goals are realized as follows. First, with the augmented query, the relevant files are obtained with the source file retrieval module. Similar as in Section~\ref{qa}, the VSM model is adopted to calculate the similarity values between the augmented query and each source file.  
Second, since we have acquired both the files retrieved by HF and the similarity values between source files and the augmented query, in the file ranking module, it is natural to combine these two types of information, to better capture the problematic files. For example, we can modify Equation~\ref{sim1} and apply $Sim^\prime$ to rank the source files:

\begin{equation}
	Sim^\prime(\vec l, \vec s) = (1 - \alpha) \times Sim(\vec l, \vec s) + \alpha \times w_s, 
	\label{sim2}
\end{equation}
where $w_s =1$ for those source files matched by the HF component, and $w_s = 0$ otherwise. $\alpha \in [0, 1]$ is a weight parameter to balance the two terms, e.g., large $\alpha$ values make RepLoc favor the HF component.

With Equation~\ref{sim2}, the source files are ranked according to their modified similarity to the augmented query, and the top ranked files are returned as the final results of RepLoc. 
We should note that, in this study, we adopt the file-level localization paradigm, in that the fixing for many unreproducible packages is not unique. For instance, statements declaring missing environmental variables can appear anywhere in the file before it is needed. Hence, it is difficult to establish line-level ground-truth. In Algorithm~\ref{algo-reploc}, we present the pseudo-code of RepLoc, which combines QA (lines 2--6), HF (lines 7--10), and FR (lines 11-16) sequentially.

\begin{table}
\caption{Files retrieved by RepLoc and its components over \texttt{dietlibc}, with successful hits in bold }
\label{example-dietlibc}
\setlength{\tabcolsep}{2pt}
\footnotesize
\begin{tabular}{|c|l|c|l|}
    \hline
Rank & \multicolumn{1}{c|}{FR (without QA)} & Rank & \multicolumn{1}{c|}{FR (with QA)} \\
    \hline
1 & /CHANGES               & 1 & \textbf{/debian/rules}            \\
2 & \textbf{/debian/rules} & 2 & \textbf{/Makefile}                \\
3 & \textbf{/Makefile}     & 3 & /CHANGES                          \\
4 & /debian/control        & 4 & /debian/patches/0005-[\dots].diff \\
5 & /FAQ                   & 5 & /diet.c                           \\
\hline
\hline
Rank & \multicolumn{1}{c|}{HF} & Rank & \multicolumn{1}{c|}{RepLoc} \\
\hline
1 & /t.c                             & 1 & \textbf{/debian/rules}        \\
2 & \textbf{/debian/implicit}        & 2 & \textbf{/Makefile}            \\
3 & /debian/dietlibc-dev.postinst.in & 3 & /CHANGES                      \\
4 & \textbf{/debian/rules}           & 4 & \textbf{/libpthread/Makefile} \\
5 & /libugly/gmtime.c                & 5 & \textbf{/libdl/Makefile}      \\
\hline
\end{tabular}
\end{table}

\textbf{Running example:} In Table~\ref{example-dietlibc}, we present the top five files retrieved by RepLoc and its individual components.  From the table, we can observe that without augmenting the query, FR is able to retrieve two problematic files. However, the topmost ranked file is a changelog (\texttt{/CHANGES}), in that the file names in the diff log appear in this file. In contrast, with the query augmented, FR (with QA) is able to rank the two problematic files at the top of the list.  
Meanwhile, although HF is able to capture \texttt{/libpthread/Makefile}, the file is not assigned top rank due to other false alarms, e.g., \texttt{/t.c}. Finally, by combining FR, QA, and HF, RepLoc is able to locate four problematic files.

\section{Experimental Results}
\label{experiment}
\subsection{Research Questions}
In this study, we intend to systematically analyze RepLoc, by investigating the following Research Questions (RQs):
\begin{itemize}[leftmargin=*]
	\item RQ1: Is RepLoc sensitive to the weighting parameter $\alpha$?
	\item RQ2: How effective is RepLoc?
	\item RQ3: How efficient is RepLoc?
	\item RQ4: Is RepLoc helpful in localizing unfixed packages?
\end{itemize}

Among these RQs, RQ1 concentrates on the impact of the weighting scheme between the components in RepLoc. RQ2 focuses on how well RepLoc performs in terms of different quality metrics. RQ3 examines whether RepLoc is time consuming, and RQ4 investigates the RepLoc's generalization. 

\subsection{Data Preparation}
\label{preparation}

In this study, the dataset is constructed by mining Debian's BTS. To the best of our knowledge, Debian is the only repository providing both past-version packages and reproducibility-related patches, which are crucial for generating the corpus and the ground truth. Consequently, all the packages within the dataset are extracted from Debian' BTS, which are tagged as unreproducible by bug reporter via \texttt{debtags}, i.e., the command line interface for accessing the BTS. 
According to Debian's documentation, there are 14 categories of reproducible issues \cite{bugsfiled}. There are also two special categories indicating the packages that fail to build from source, and the tool-chain issues (non-deterministic issues introduced by other packages, see Section~\ref{threats}), which are not considered in this study. 

\begin{figure}[t]
	\centering
	\subfigure[Number of packages]{
	\includegraphics[width=.28 \textwidth]{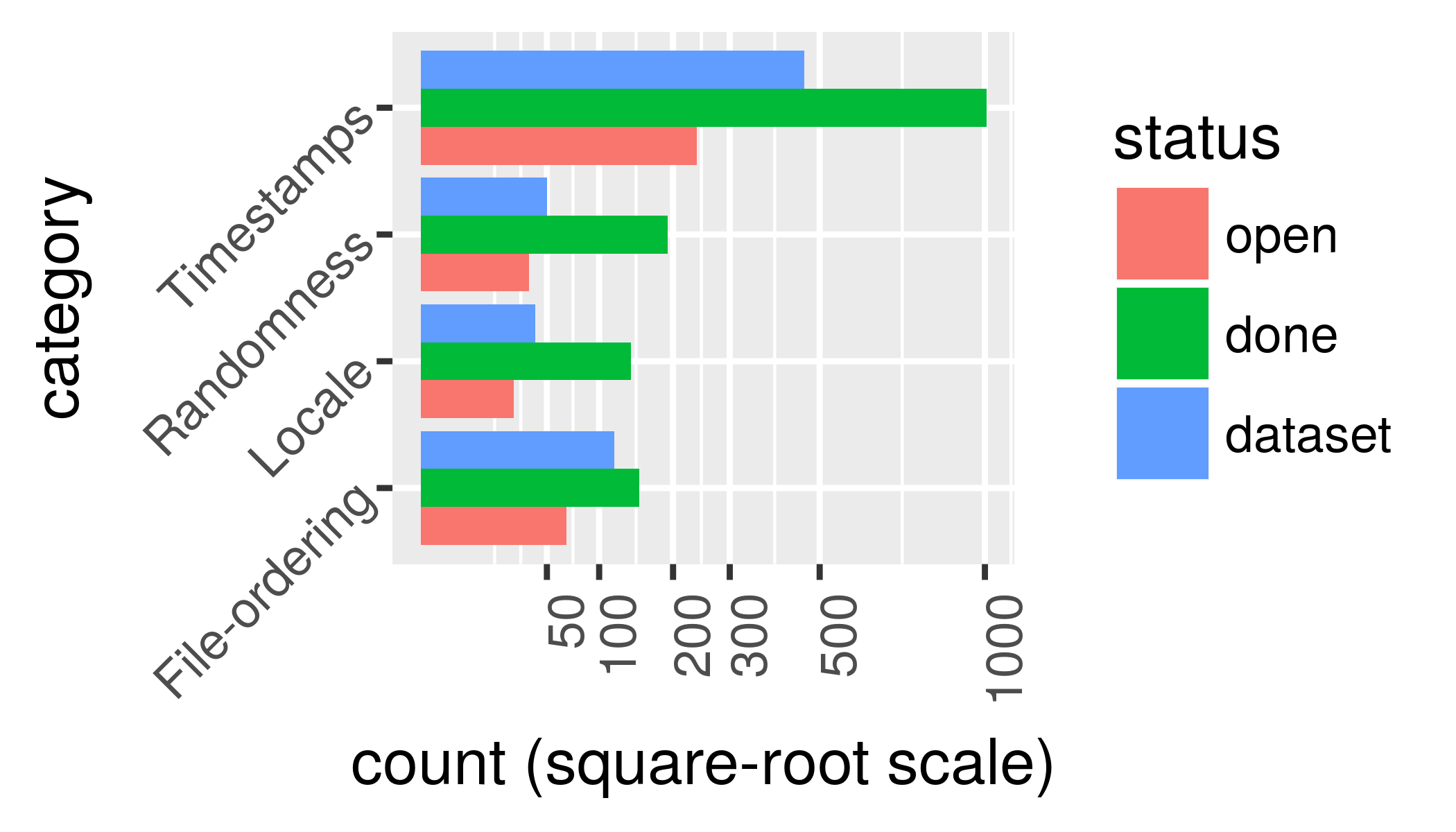}
	\label{summary}
	}
	\subfigure[Problematic file types]{
		\raisebox{.17 in}{
			\includegraphics[width=.16 \textwidth]{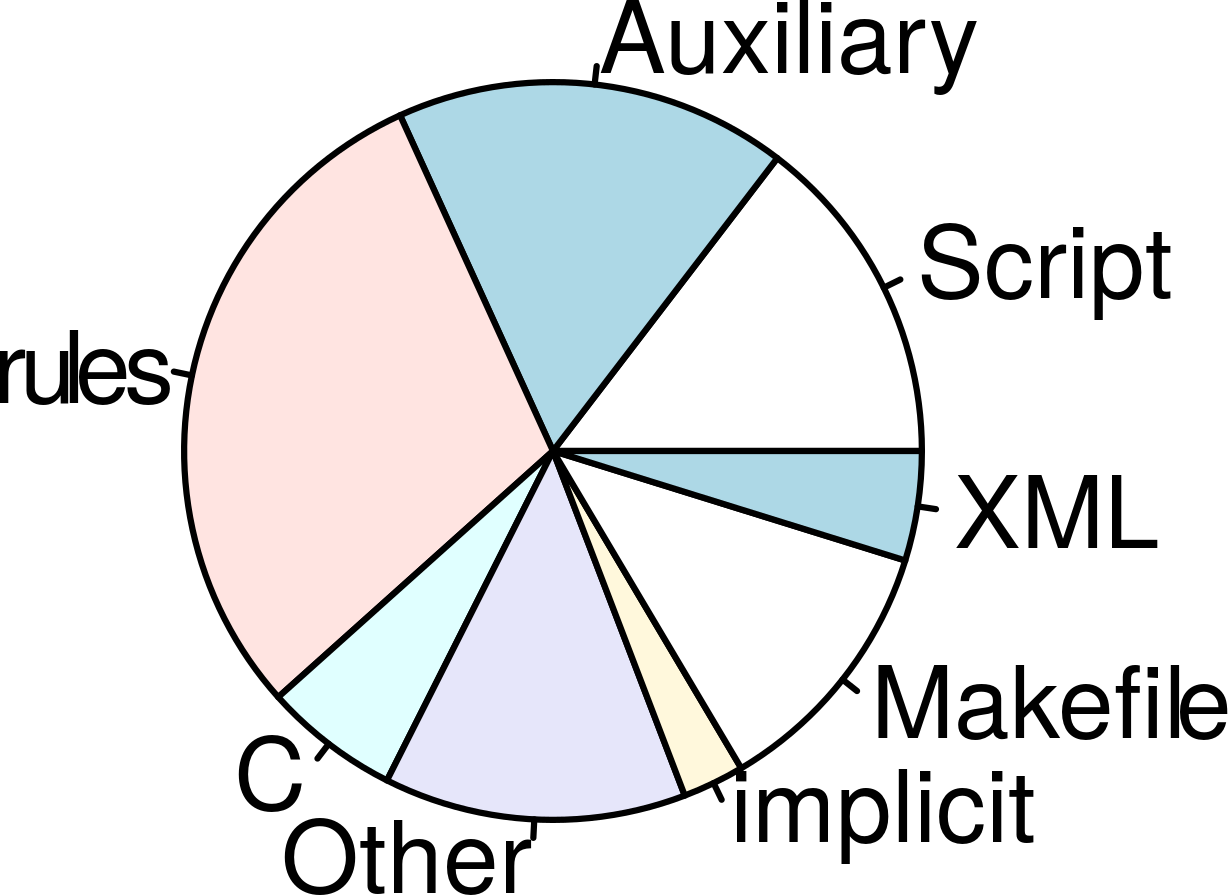}
		}
	\label{pie}
	}
	\caption{File statistics of the dataset}
\end{figure}

We download all the 14 categories of 1716 bug reports, and download the packages, with their corresponding patches. Then, we apply the validation tool kit,\footnote{The tool kit realizes steps 1--3 of Figure~\ref{flow}, available at \url{https://anonscm.debian.org/cgit/reproducible/misc.git}} to obtain the corresponding diff logs and build logs. In this study, we consider those categories with more than 30 packages. With such criterion, we obtain 671 packages in the dataset, which fall into the four largest categories.  
Figure~\ref{summary} illustrates the statistics of the dataset. In the figure, we present the numbers of the open and closed bugs in Debian's BTS, as well as the number of packages in the dataset. Among the four categories of packages, the Timestamps category contains the most packages (462), followed by File-ordering (118), Randomness (50), and Locale (41). For all the four categories of 1491 packages that are labeled as ``done'', the packages in the dataset take a portion of $45.34\%$. Note that there are less packages in the dataset than closed bug reports, since packages may not be compilable due to the upgrade of their dependencies. 

In Figure~\ref{pie}, we illustrate the statistics of the patches in the dataset.  
From the figure, we could observe that there are many types of files that might be involved in the unreproducible builds. For these files, the Debian \texttt{rules} files, which are the main build scripts, take the largest portion of the fixed files (29.82\%). Auxiliary files, such as the \texttt{configure} scripts and  input files (\texttt{*.in}), takes the second largest portion (17.21\%). After that, there are the Makefiles (11.68\%), scripts such as Python/Perl/PHP files (14.60\%), C/C++ files  (5.94\%), XML files (4.80\%), implicit build files (2.71\%).
Since we classify the files based on their file extensions heuristically, there are also 13.24\% of the files that are not easy to classify, e.g, those without file extensions. 
This phenomenon conforms with the second barrier mentioned in Section~\ref{intro}, i.e., the causes to the unreproducible builds are diverse, which makes the localization task very challenging.

\subsection{Implementation and Metrics}
\label{metrics}
RepLoc is implemented in Perl 5.24 and Java 1.8. All the experiments are conducted on an Intel Core i7 4.20 GHz CPU server with 16 GB memory, running GNU/Linux with kernel 4.9.0.  
For the comparative algorithms, we consider four variants of RepLoc, since there is no prior approach addressing this problem.  The first two variants implement two baseline algorithms, which only consider either the HF or the FR model (denoted as RepLoc(HF) and RepLoc(FR)). These two variants are incorporated to examine the performance of its building-block components. Moreover, RepLoc(FR) could be considered the simulation of the manual localization, since in FR, the retrieval is realized by matching source files with diff log contents. Then, RepLoc(FR+QA) considers utilizing the QA component to enhance the basic queries extracted from the diff logs. Finally, RepLoc indicates the version discussed in Section~\ref{framework}.

To evaluate the effectiveness of RepLoc, metrics commonly used in the IR literatures are employed to evaluate the performance of RepLoc, including the accuracy rate, the precision, the recall, and the Mean Average Precision (\textit{MAP}). The metrics are computed by examining the ranked list of source files returned by the framework in response to a query. The Top-$N$ source files in the ranked list is called the retrieved set and is compared with the relevance list to compute the Precision and Recall metrics (denoted by \textit{P@N} and \textit{R@N} respectively). Given an unreproducible package with problematic files, a Top-$N$ accuracy rate score, e.g. \textit{A@1}, \textit{A@5}, and \textit{A@10}, of a localization tool is the portion of Top-$N$ lists a tool provides that at least one problematic file contains in it \cite{li2014learning, ye2014learning}. In this study, we also report \textit{P@1}, \textit{P@5}, \textit{P@10} and \textit{R@1}, \textit{R@5}, \textit{R@10} \cite{kochhar2014potential, ye2014learning}. \textit{P@N} means the portion of problematic files successfully retrieved in a Top-$N$ list, while \textit{R@N} measures how many problematic files are retrieved in a Top-$N$ list among all the problematic files:
 
\begin{equation}
	P@N = \frac{\text{\# of files that cause unreproducible builds}}{N},
	\label{pk}
\end{equation}
\begin{equation}
    R@N = \frac{\text{\# retrieved problematic files in the Top-}N \text{ list}}{\text{\# of problematic files}}.
	\label{rk}
\end{equation}

\begin{table*}[t]
	\centering
    \caption{Comparison results between RepLoc and its variants}
	\label{effectiveness-tfidf}
	\footnotesize
\begin{tabular}{|l|l|c|c|c|c|c|c|c|c|c|c|}
  \hline
 Dataset & Model & \textit{A@1} & \textit{A@5} & \textit{A@10} & \textit{P@1} & \textit{P@5} & \textit{P@10} & \textit{R@1} & \textit{R@5} & \textit{R@10} & \textit{MAP} \\ 
	\hline
	\multirow{4}{*}{Timestamps}
    & RepLoc(HF)    & 0.4048 & 0.6775 & 0.7229 & 0.4048 & 0.1511 & 0.0835 & 0.3587 & 0.6222 & 0.6682 & 0.3522 \\ 
    & RepLoc(FR)    & 0.3160 & 0.5736 & 0.7121 & 0.3160 & 0.1268 & 0.0807 & 0.2821 & 0.5253 & 0.6553 & 0.2777 \\ 
    & RepLoc(FR+QA) & 0.4762 & 0.6753 & 0.7641 & 0.4762 & 0.1511 & 0.0883 & 0.4155 & 0.6177 & 0.7102 & 0.4009 \\ 
    & RepLoc        & \textbf{0.5238} & \textbf{0.7792} & \textbf{0.8290} & \textbf{0.5238} & \textbf{0.1792} & \textbf{0.0991} & \textbf{0.4538} & \textbf{0.7295} & \textbf{0.7839} & \textbf{0.4400} \\ 
	\hline
	\multirow{4}{*}{File-ordering}
    & RepLoc(HF)    & 0.3136 & 0.4407 & 0.4576 & 0.3136 & 0.0983 & 0.0534 & 0.2653 & 0.3968 & 0.4197 & 0.2528 \\ 
    & RepLoc(FR)    & 0.1525 & 0.5169 & 0.6949 & 0.1525 & 0.1085 & 0.0729 & 0.1215 & 0.4427 & 0.6150 & 0.1136 \\ 
    & RepLoc(FR+QA) & 0.3814 & 0.6780 & 0.7627 & 0.3814 & 0.1492 & 0.0864 & 0.3040 & 0.5978 & 0.6856 & 0.2804 \\ 
    & RepLoc        & \textbf{0.4492} & \textbf{0.7288} & \textbf{0.7966} & \textbf{0.4492} & \textbf{0.1661} & \textbf{0.0966} & \textbf{0.3774} & \textbf{0.6506} & \textbf{0.7331} & \textbf{0.3572} \\ 
	\hline
	\multirow{4}{*}{Randomness}
    & RepLoc(HF)    & 0.1000 & 0.2200 & 0.2600 & 0.1000 & 0.0480 & 0.0280 & 0.0850 & 0.2100 & 0.2500 & 0.0813 \\ 
    & RepLoc(FR)    & 0.1000 & 0.3000 & 0.4800 & 0.1000 & 0.0640 & 0.0500 & 0.1000 & 0.3000 & 0.4650 & 0.1000 \\ 
    & RepLoc(FR+QA) & \textbf{0.2200} & 0.3200 & 0.4200 & \textbf{0.2200} & 0.0680 & 0.0460 & \textbf{0.2100} & 0.3050 & 0.4100 & \textbf{0.2050} \\ 
    & RepLoc        & 0.2000 & \textbf{0.4200} & \textbf{0.5000} & 0.2000 & \textbf{0.0880} & \textbf{0.0540} & 0.1900 & \textbf{0.4050} & \textbf{0.4900} & 0.1854 \\ 
	\hline
	\multirow{4}{*}{Locale}
    & RepLoc(HF)    & 0.0976 & 0.3171 & 0.3659 & 0.0976 & 0.0634 & 0.0366 & 0.0976 & 0.3049 & 0.3415 & 0.0976 \\ 
    & RepLoc(FR)    & 0.1463 & 0.2439 & 0.4634 & 0.1463 & 0.0488 & 0.0463 & 0.1463 & 0.2317 & 0.4512 & 0.1494 \\ 
    & RepLoc(FR+QA) & 0.2439 & 0.4146 & 0.5610 & 0.2439 & 0.0829 & 0.0561 & 0.2317 & 0.4024 & 0.5488 & 0.2256 \\ 
    & RepLoc        & \textbf{0.2683} & \textbf{0.5122} & \textbf{0.7317} & \textbf{0.2683} & \textbf{0.1024} & \textbf{0.0732} & \textbf{0.2561} & \textbf{0.5000} & \textbf{0.7195} & \textbf{0.2500} \\ 
	\hline
    \hline
	\multirow{4}{*}{Overall}
    & RepLoc(HF)    & 0.3472 & 0.5797 & 0.6200 & 0.3472 & 0.1288 & 0.0712 & 0.3059 & 0.5324 & 0.5734 & 0.2990 \\ 
    & RepLoc(FR)    & 0.2608 & 0.5231 & 0.6766 & 0.2608 & 0.1142 & 0.0750 & 0.2320 & 0.4760 & 0.6216 & 0.2278 \\ 
    & RepLoc(FR+QA) & 0.4262 & 0.6334 & 0.7258 & 0.4262 & 0.1404 & 0.0829 & 0.3694 & 0.5777 & 0.6736 & 0.3544 \\ 
    & RepLoc        & \textbf{0.4709} & \textbf{0.7273} & \textbf{0.7928} & \textbf{0.4709} & \textbf{0.1654} & \textbf{0.0937} & \textbf{0.4087} & \textbf{0.6774} & \textbf{0.7491} & \textbf{0.3949} \\ 
  \hline
\end{tabular}
\end{table*}
	
\begin{table}[ht]
\centering
	\caption{Result of RepLoc(HF), with single heuristic rule}
	\label{single-rule}
\footnotesize
\begin{tabular}{|l|l|c|c|c|c|}
  \hline
	ID & Rule & \textit{A@10} & \textit{P@10} & \textit{R@10} & \textit{MAP} \\ 
  \hline
  3  & \texttt{GZIP\_ARG}           &  0.2981 & 0.0341 & 0.2823 & 0.1864 \\
  4  & \texttt{DATE\_CMD}           &  0.2191 & 0.0253 & 0.1878 & 0.1250 \\
  14 & \texttt{UNSORTED\_WILDCARD}  &  0.1058 & 0.0112 & 0.0968 & 0.0578 \\
  13 & \texttt{LS\_WITHOUT\_LOCALE} &  0.0671 & 0.0072 & 0.0428 & 0.0247 \\
  9  & \texttt{SORT\_IN\_PIPE}      &  0.0387 & 0.0039 & 0.0351 & 0.0261 \\
   \hline
\end{tabular}
\end{table}

Precision and Recall usually share an inverse relationship, in that, the Precision is higher than Recall for lower values of $N$ and vice versa for higher values of $N$. An overall metric of retrieval accuracy is known as Mean Average Precision (\textit{MAP}), which is the average of the Average Precision (\textit{AP}) values over all the problematic files in unreproducible packages. 
For an unreproducible package with several problematic files, the \textit{AP} is computed as 
	 $\sum _{k=1}^M \frac{P@k \times pos(k)}{\text{\# of files related in the patch}}$,
 where $M$ is the size of a ranking list, $pos(k)$ indicates whether the $k$\textit{th} file in a ranking list is related to the unreproducible build, and $P@k$ is the precision described in Equation~\ref{pk}. With \textit{AP} defined, \textit{MAP} can be calculated by averaging all the \textit{AP} scores across all the unreproducible packages.

\subsection{Investigation of RQ1}
\label{rq1}

\begin{figure}[t]
	\centering
	\includegraphics[width=.45 \textwidth]{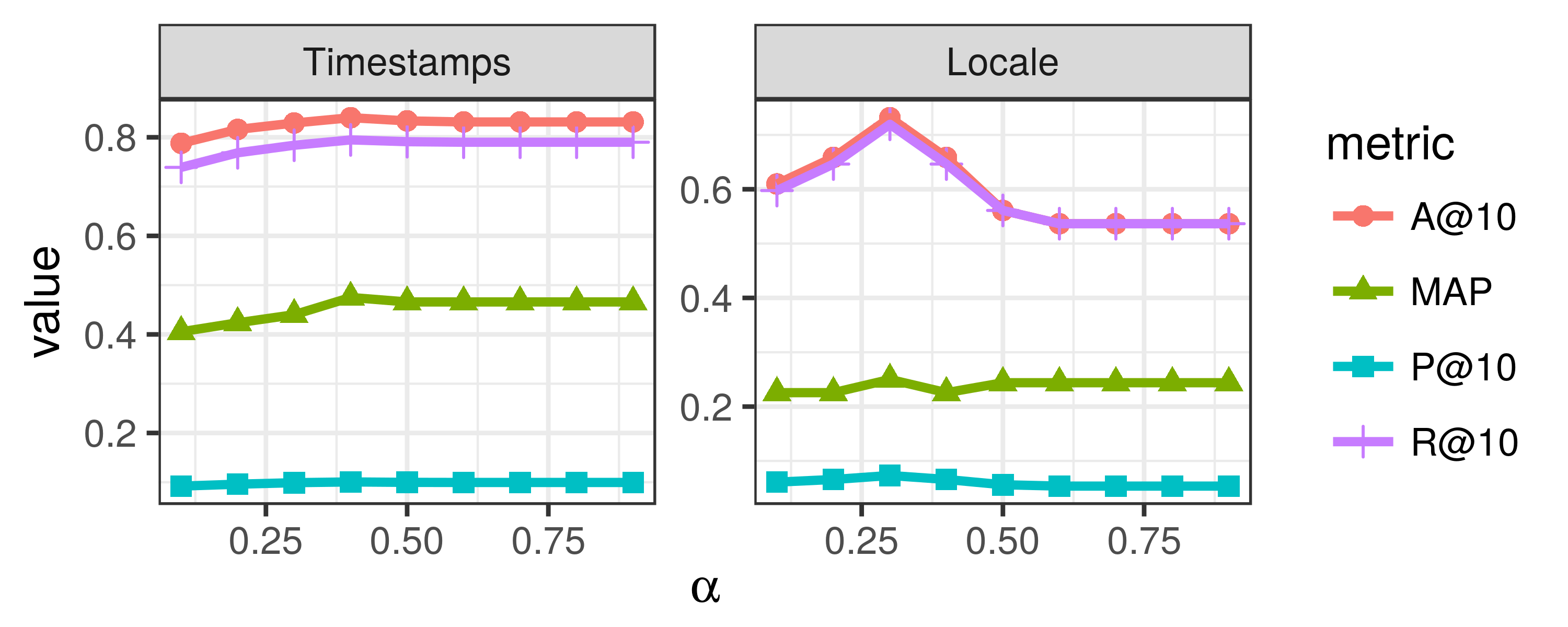}
	\caption{Impact of varying $\alpha$}
	\label{alpha}
\end{figure}

In this RQ, we intend to investigate whether RepLoc is sensitive to the weighting parameter $\alpha$. As described in Section~\ref{framework}, in Equation~\ref{sim2}, we propose the weighted similarity between queries and source files. Hence, in this RQ, we are interested in investigating RepLoc's behavior as we alter the weight of the two components. More specifically, for each category of dataset, we randomly select half of the packages, and a grid search from 0.1 to 0.9 with a step of 0.1 is employed to analyze the impact of varying $\alpha$.

Considering the Timestamps and the Locale datasets, we visually present the trend of the \textit{A@10}, \textit{P@10}, \textit{R@10} and the \textit{MAP} values against the $\alpha$ value in Figure~\ref{alpha}. From the figure, the following observations can be drawn. First, for the randomly selected packages, the performance of RepLoc exhibits similar trend, i.e., when $\alpha$ is set within the range $[0.2, 0.4]$, RepLoc obtains the best results. Second, we observe that RepLoc is not very sensitive to $\alpha$, unless $\alpha$ is too large, which will make RepLoc prefer the HF component. Hence, for the subsequent experiments, $\alpha$ is set with 0.3. 

\textbf{Answer to RQ1:} Experimental results show that, RepLoc is not very sensitive to the parameter, which to some extent demonstrates the robustness of RepLoc.

\subsection{Investigation of RQ2}
\label{rq2}

In this RQ, we examine whether RepLoc locates the problematic files accurately. We present the experimental results, and discuss the phenomena observed. In Table~\ref{effectiveness-tfidf}, we first give the results over the datasets. The table is organized as follows. The first column indicates the four categories of datasets we built in this study (see Section~\ref{preparation}). The second column represents the four variants of RepLoc. Then, the rest of the table presents the metrics that evaluate the performance of each variant. Note that for the accuracy rate, the precision, and the recall, the metric values are averaged over all the packages. Besides, we also present the aggregate performance at the bottom of the table.

Taking the Timestamps dataset as an example, several interesting phenomena can be observed. First, the performance of RepLoc(HF) is not satisfying. 
Even considering the Top-10 results, the corresponding accuracy rate is around $70\%$. To examine the representativeness of the heuristic rules, in Table~\ref{single-rule} we present the results of RepLoc(HF) with single rule. We report the \textit{A@10}, \textit{P@10}, \textit{R@10}, and \textit{MAP} of the five rules that perform the best. Among the rules, the \texttt{GZIP\_ARG} rule achieves the highest accuracy rate. However, the  \textit{A@10} value is below $30\%$, which is significantly outperformed by RepLoc(HF) that considers all the rules. Similar observations could be drawn for other performance metrics, which to some extent confirms the diverse-cause barrier. 

Second, by comparing the results of RepLoc(FR+QA) against RepLoc(FR) in Table~\ref{effectiveness-tfidf}, we can confirm the usefulness of QA. As mentioned, RepLoc(FR) could be loosely considered the simulation of manual localization, which tries to match the problematic files with the diff log contents. Over the Timestamps dataset, \textit{A@10} of RepLoc(FR) is $71.21\%$. With the augmentation of the query, \textit{A@10} improves to $76.41\%$. Moreover, when we combine RepLoc(FR+QA) with HF, the performance is further improved, i.e., \textit{A@10} of RepLoc achieves $82.90\%$, which implies that for over $80\%$ of the unreproducible packages in the Timestamps dataset, at least one problematic file is located in the Top-10 list. Besides, similar results are obtained over the other datasets, i.e., RepLoc(HF) and RepLoc(FR) perform the worst, RepLoc(FR+QA) outperforms RepLoc(FR) considering the \textit{A@10} value, and RepLoc performs the best.

Associated with Table~\ref{effectiveness-tfidf}, we also conduct statistical tests, to draw confident conclusions whether one algorithm outperforms the other. For the statistical test, we employ the Wilcoxon's signed rank test, with a null hypothesis stating that there exists no difference between the results of the algorithms in comparison. We consider the $95\%$ confidence level (i.e., \textit{p}-values below 0.05 are considered statistically significant), and adopt the \textit{P@10} and \textit{R@10} as the performance metrics. We do not consider the accuracy rate and the MAP metrics, in that these are aggregate metrics.
Over all the instances, when comparing RepLoc with any of the other three baseline variants, the null hypothesis is rejected (\textit{p}-value $ < 0.05$ for both \textit{P@10} and \textit{R@10}), which implies that RepLoc outperforms their baseline variants in a statistically significant way.

\begin{figure}[t]
	\centering
	\includegraphics[width=.4 \textwidth]{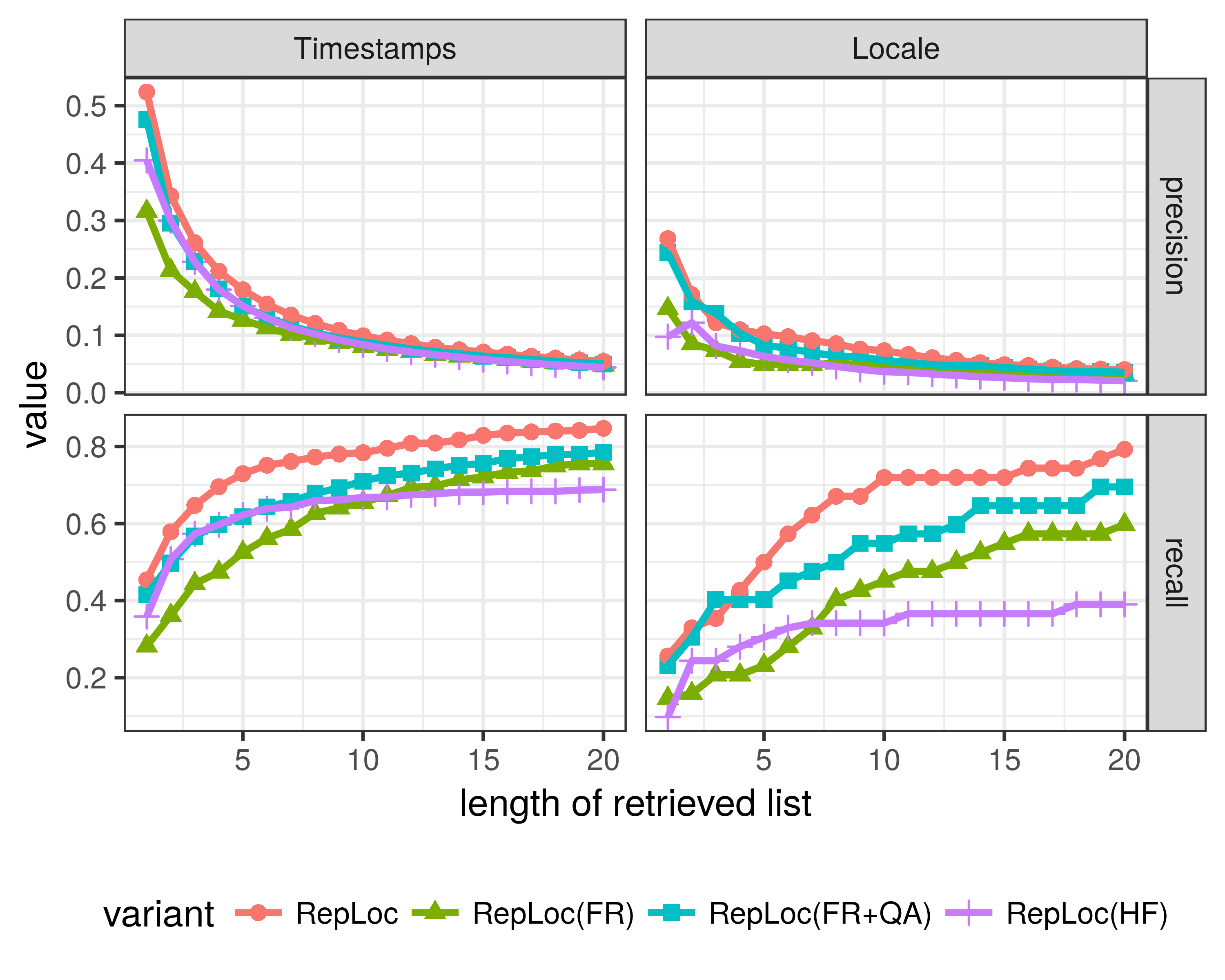}
	\caption{Trends of precision and recall of RepLoc}
	\label{trend}
\end{figure}

To gain more insights into the behavior of RepLoc, we present the performance of the four variants against the number of retrieved results in Figure~\ref{trend}, over typical datasets. In the figure, the x-axis and the y-axis indicate the number of retrieved files, and the performance metrics. From the sub-figures, we confirm that over both the datasets, RepLoc outperforms the other variants significantly, i.e., the performance curves for RepLoc lie above those for other variants, which implies that for all the cases of the retrieved results, combining the two components is able to obtain better results. This phenomenon conforms with our observations in Table~\ref{effectiveness-tfidf}. 

\textbf{Answer to RQ2:} By comparing the variants of RepLoc over 671 real world packages, we confirm that by combining the heuristic rule-based filter and the query augmentation, RepLoc is able to outperform its variants.

\subsection{Investigation of RQ3}
\label{rq3}
\begin{figure}[t]
	\centering
	\includegraphics[width=.45 \textwidth]{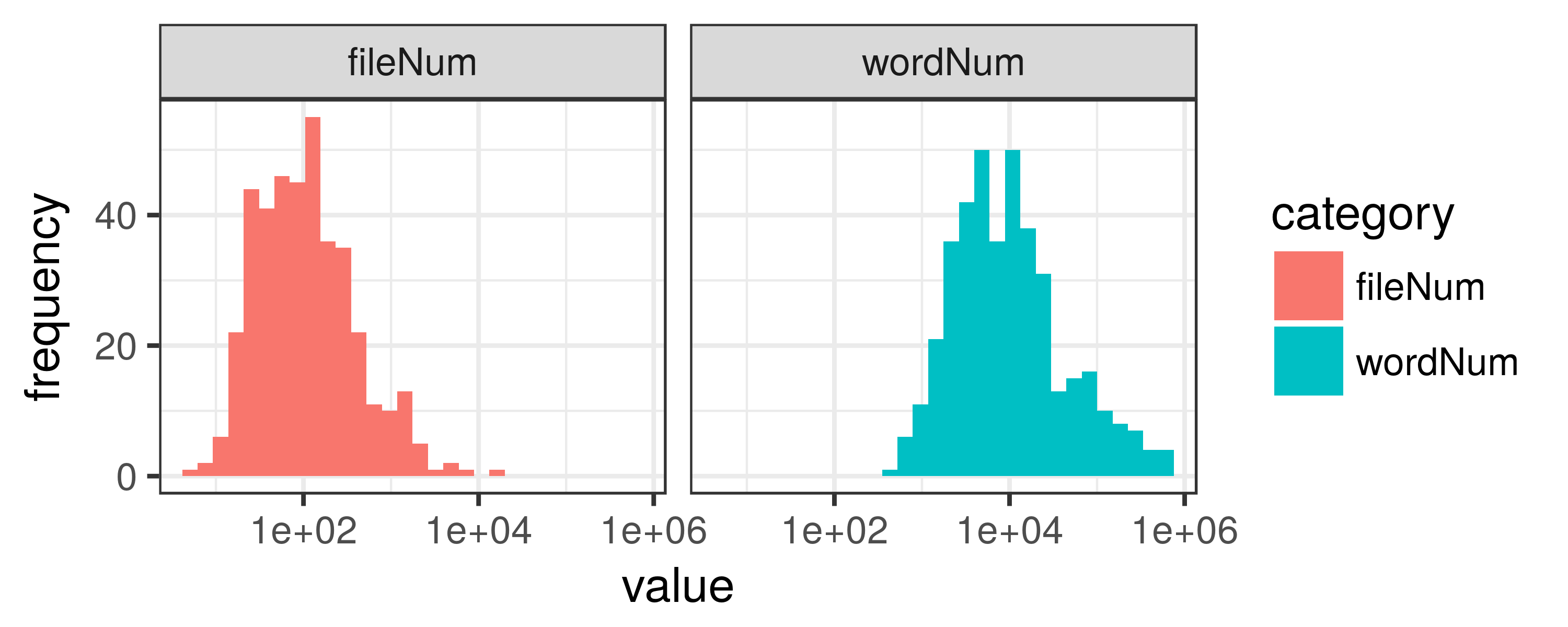}
	\caption{Histogram for scale statistics}
	\label{scale-stat}
\end{figure}

In this RQ, we evaluate RepLoc from the efficiency perspectives. Since manually localizing the unreproducible issues is a time consuming task, automating such process is profitable only if the proposed approach is time efficient. Hence, we present the time statistics of the experiments.  
Figure~\ref{scale-stat} depicts the statistics of the source files as histograms, in which the x-axis indicates the number of source files (fileNum) and the words (wordNum), and the y-axis represents the associated frequency. In this study, the number of files ranges within [6, 19890], and the number of words for the majority of the packages ranges around $1 \times 10^4$, which implies that manually inspecting the files would be difficult.

\begin{figure}[t]
	\centering
	\includegraphics[width=.5 \textwidth]{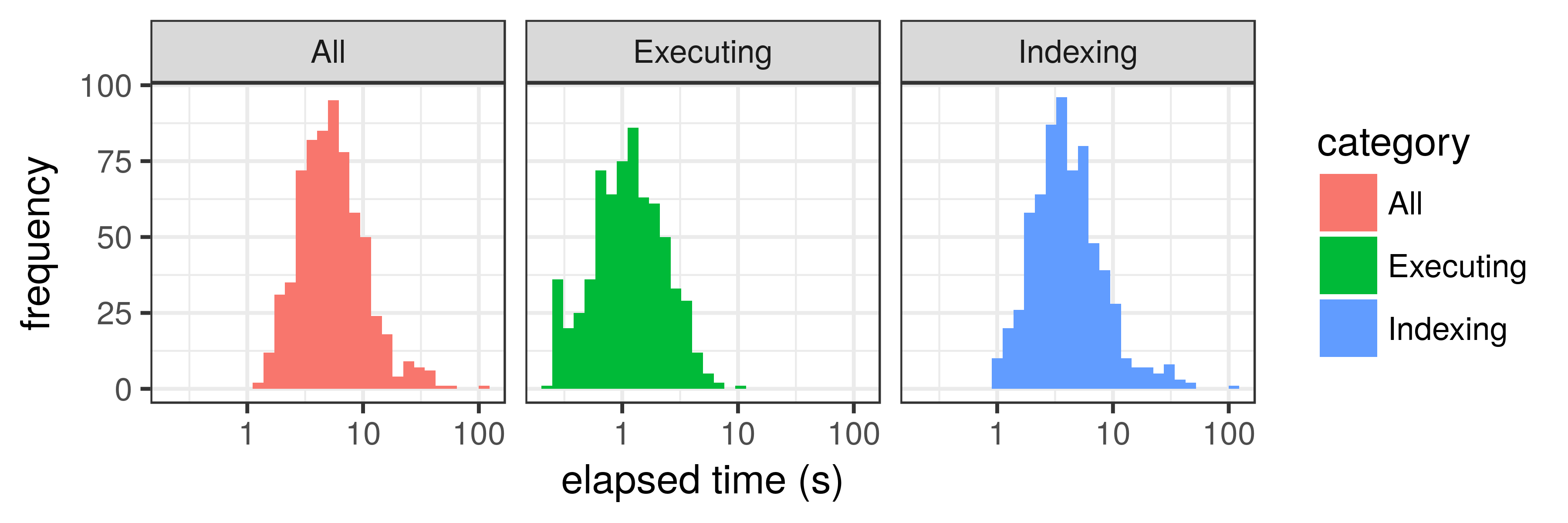}
	\caption{Histogram for efficiency evaluation}
	\label{time-stat}
\end{figure}

Since the scale of the packages in this study varies greatly, it is intuitive that the localization process over different packages will vary accordingly. To investigate this issue, we present the results related to time efficiency considering the three variants of RepLoc. 
In Figure~\ref{time-stat}, we illustrate the distributions of the dataset scalability and the execution time. 
In the sub-figures, the x-axis indicates the time in seconds, and the y-axis represents the frequency.  From the results, we observe that, the indexing of the documents consumes the largest portion of time, compared with other components. In particular, the median of the execution time for RepLoc is 5.14 seconds. 

\textbf{Answer to RQ3:} In this RQ, we investigate the efficiency perspectives of RepLoc. In this study, the indexing of the document consume the majority of the time.

\subsection{Investigation of RQ4}
\label{rq4}

For RQ1--RQ3, to evaluate the performance of RepLoc properly, we employ the packages that have been fixed, and adopt the patches from the BTS as the ground truth. However, in the real-world reproducible validation scenario, the patches are not available in advance. Hence, in this RQ, we intend to investigate RepLoc under such condition. More specifically, we consider two scenarios, i.e., we apply RepLoc to the packages over (1) Debian packages that are previously unfixed, and (2) the unreproducible packages from Guix.

First, we are interested in whether RepLoc could be generalized to unfixed packages, which are obtained from the continuous integration system of Debian. We also check the BTS, to ensure that the packages have not been fixed. We apply RepLoc to localize the problematic files, and then manually check and fix the unreproducible issues. Through localization and fixing, 3 unreproducible packages belonging to the Timestamps category are fixed, i.e., \texttt{regina-rexx} (3.6-2), \texttt{fonts-uralic} (0.0.20040829-5), and \texttt{manpages-tr} (1.0.5.1-2). We submit the corresponding patches to the BTS \cite{bugreportfontsuralic, bugreportmanpagestr, bugreportregina}, and the one for \texttt{fonts-uralic} has been accepted. 

\begin{table}[t]
	\caption{Results of RepLoc, over \texttt{manpages-tr}}
	\label{example-manpages-tr}
	\setlength{\tabcolsep}{2pt}
	\centering 
	\footnotesize
	\begin{tabular}{|c|l|c|l|}
		\hline
Rank & \multicolumn{1}{c|}{RepLoc(FR)} & Rank & \multicolumn{1}{c|}{RepLoc(FR+QA)} \\
		\hline
1 & /debian/rules                & 1 & /debian/patches/bashisms.patch \\
2 & /source/man8/mount.8.xml     & 2 & /debian/rules                  \\
3 & /source/tr/linkata.sh        & 3 & \textbf{/source/manderle.sh}   \\
4 & /source/man1/rsync.1.xml     & 4 & /Makefile                      \\
5 & \textbf{/source/manderle.sh} & 5 & /debian/manpages-tr.prune      \\
\hline
\hline
Rank & \multicolumn{1}{c|}{RepLoc(HF)} & Rank & \multicolumn{1}{c|}{RepLoc} \\
		\hline
1 & /source/man1/gzip.1.xml      & 1 & \textbf{/source/manderle.sh}   \\
2 & \textbf{/source/manderle.sh} & 2 & /debian/patches/bashisms.patch \\
3 &                              & 3 & /debian/rules                  \\
4 &                              & 4 & /source/man1/gzip.1.xml        \\
5 &                              & 5 & /source/man1/patch.1.xml       \\
\hline
	\end{tabular}
\end{table}

For these packages, the problematic files are ranked among the top of the retrieved list by RepLoc. For example, in Table~\ref{example-manpages-tr}, we present the results over the package \texttt{manpages-tr}. The table is organized similarly as Table.~\ref{example-dietlibc}. From the table, we observe that RepLoc is able to localize problematic files effectively, i.e., the problematic files are ranked the first in the result. The package is unreproducible due to the invocation of \texttt{gzip} without ``\texttt{-n}'', and the issue can be captured by the \texttt{GZIP\_ARG} rule in ``\texttt{/source/manderle.sh}''. However, since the heuristic rules fail to capture the dynamic aspect of the build process, a file (``\texttt{/source/man1/gzip.1.xml}'') unused during compilation is also retrieved. In contrast, with FR and QA, we concentrate on the files involved by the build process. By combining both the static (HF) and the dynamic (HF and QA) perspectives, the problematic file is ranked the first of the list with higher probability. 

Second, we consider the packages from the Guix repository, to investigate whether the knowledge obtained from Debian could be generalized to other repositories. The reasons we choose Guix are that, (1) the repository is interested in the reproducible builds practice \cite{guixrepro}, and (2) its package manager provides the functionality of validating package reproducibility locally, which facilitates the experimental design. As a demonstration, we localize and manually fix the problematic files of 3 packages, namely \texttt{libjpeg-turbo} (1.5.2), \texttt{djvulibre} (3.5.27), and \texttt{skalibs} (2.3.10.0). Similar with the previous case, the patches were submitted to Guix's BTS \cite{bugreportlibjpeg-turbo,bugreportdjvulibre,bugreportskalibs}. Taking \texttt{skalibs} as an example, we present the results of the variants of RepLoc in Table~\ref{example-guix}. From the table, we could observe that the problematic file ``\texttt{/Makefile}'' is assigned the top rank. Contrarily, without RepLoc, over 900 source files have to be manually traversed. Such observation to some extent demonstrates the usefulness of RepLoc in leveraging the knowledge from Debian to a different repository such as Guix. After localizing the problematic file and manually fixing, the submitted patch has been accepted and pushed into the code base of Guix \cite{bugreportskalibs}. Similarly, the patches for \texttt{djvulibre} \cite{bugreportdjvulibre} and \texttt{libjpeg-turbo} \cite{bugreportlibjpeg-turbo} have also been accepted. 

\begin{table}[t]
	\caption{Results of RepLoc, over \texttt{skalibs}}
	\label{example-guix}
	\setlength{\tabcolsep}{2pt}
	\centering 
	\footnotesize
	\begin{tabular}{|c|l|c|l|}
		\hline
Rank & \multicolumn{1}{c|}{RepLoc(FR)} & Rank& \multicolumn{1}{c|}{RepLoc(FR+QA)} \\
		\hline
1 & /package/info                              & 1 & /configure\\
2 & /doc/[\dots]/kolbak.html             & 2 & /src/[\dots]/uint32\_reverse.c\\
3 & /doc/[\dots]/unixmessage.html        & 3 & /src/[\dots]/badrandom\_here.c\\
4 & /doc/[\dots]/unix-transactional.html & 4 & /src/[\dots]/goodrandom\_here.c\\
5 & /doc/[\dots]/unix-timed.html         & 5 & /src/[\dots]/md5\_transform.c\\
\dots & \dots& \dots&\dots\\
24 & \textbf{/Makefile}                                  & 10 & \textbf{/Makefile}            \\
\hline
\hline
Rank & \multicolumn{1}{c|}{RepLoc(HF)} & Rank & \multicolumn{1}{c|}{RepLoc} \\
		\hline
1 & /tools/gen-deps.sh                    & 1 & \textbf{/Makefile}                       \\
2 & \textbf{/Makefile}                    & 2 & /configure                      \\
3 & /src/[\dots]/localtm\_from\_ltm64.c & 3 & /src/[\dots]/uint32\_reverse.c \\
4 &                                       & 4 & /src/[\dots]/badrandom\_here.c \\
5 &                                       & 5 & /src/[\dots]/goodrandom\_here.c\\
\hline
	\end{tabular}
\end{table}

\textbf{Answer to RQ4:} We demonstrate that RepLoc is helpful in localizing unfixed unreproducible packages from both Debian and Guix. In particular, unreproducible issues of 6 packages from both repositories are fixed under the guidance of RepLoc, which have not been fixed before this study.

\section{Threats to Validity}
\label{threats}

There are several objections a critical reader might raise to the evaluation presented in this study, among which the following two threats deserve special attention. 

First, in this study, the heuristic rules in HF are summarized from Debian's documentation. Also, we leverage the build log gathered from the build process. Hence, some may argue that the approach cannot be generalized to other software repositories because it relies too much on Debian's infrastructure. To mitigate this threat, in RepLoc, attention is paid so that the components are not specialized for Debian. For example, despite knowing that the Debian \texttt{rules} files take the largest portion of the problematic files (see Figure~\ref{pie}), no extra priority is given to these files during ranking. Also, in HF, we avoid using heuristic rules specific to Debian, and intend to make the rules as general as possible. For instance, \texttt{UNSORTED\_WILDCARD} is applicable for Makefile based build systems, and \texttt{GZIP\_ARG} is helpful if \texttt{gzip}-based compression is involved. As a result, the results of this study can be generalized to other repositories. As demonstrated in RQ4, we have successfully applied RepLoc to Guix. For other repositories, applying RepLoc should only require minor adaptation. For example, for the Fedora project, the build log can be gathered by parsing the verbose output of the \texttt{mock} build tool, and the diff log could be generated by \texttt{diffoscope} as well.

Second, when constructing the datasets, the unreproducible packages caused by the tool-chain issues are not considered. For these packages, the unreproducible issues are introduced by the depended packages rather than the current package. Hence, identification of the tool-chain issues is another challenging task that requires further manual investigation \cite{fixtoolchain}. Besides, we should note that fixing the tool-chain issues may help make more packages reproducible. For example, when reproducible-related patches were accepted by \texttt{gcc} from upstream, around 200 unreproducible packages that depended on \texttt{gcc} became reproducible automatically \cite{weeklyreport}. 
We plan to explore the tool-chain issues in the future.
\section{Related Work}
\label{related}
\subsection{Bug Localization Related Work}
First, this study is closely related to the fault localization studies, especially the IR-based approaches. 

For example, Zhou et al. \cite{zhou2012should} proposed a specialized VSM based approach, and consider the similarities between bug reports to localize buggy files.  Wang et al. \cite{wang2014compositional} propose a compositional model that integrates multiple variants of VSM. In particular, they model the composition of different VSM variants as a optimization problem, and apply a genetic algorithm to search for the suitable composition pattern between VSM variants.  Wang et al. \cite{Wang:2015:EUI:2771783.2771797} investigate the usefulness of IR-based fault localization techniques, and discover that the quality of the bug reports are crucial to the performance of localization tasks. 

Meanwhile, domain knowledge is utilized to improve the performance of IR-based bug localization techniques. Ye et al. \cite{ye2014learning} find bug-fixing frequency and bug-fixing recency of source code files are helpful for bug localization. Saha et al. \cite{saha2013improving} find the structure of bug reports and source code files are also good knowledge for bug localization. They consider bug reports or source code files as documents with structured fields, e.g., summary and description, or file name, class name, and method name, respectively. 
Stack-trace information in bug report is also analyzed \cite{wong2014boosting, DBLP:conf/icst/MoranVBVP16} to improve the performance of bug localization. Besides, version histories \cite{wang2014version, sisman2012incorporating, tantithamthavorn2013using} and similar bug reports \cite{davies2012using} are proved to be useful. 

Besides, with the development of IR techniques, other text mining methodologies are also incorporated to support locating buggy files.  For example, due to its effectiveness, Latent Dirichlet Allocation (LDA) has gained its popularity in the field of bug localization.  Lukins et al. \cite{lukins2010bug} propose a static LDA-based technique for automatic bug localization.  Lam et al. \cite{lam2015combining} propose a localization framework HyLoc that combines deep learning and IR-based model. They integrate deep neural network and a VSM variant, to complement the two standalone components. Experimental results over real world projects demonstrate that their proposed model outperforms the individual models.
Rao et al. \cite{rao2015comparing} propose an  incremental framework to update the model parameters of the Latent Semantic Analysis, which is then applied to localize buggy files. Experiments over software libraries with ten years of version history validate their framework.

However, despite the closeness to these studies, we should note that the problem in this study has its unique features. For example, the counterpart of the bug reports in IR-based fault localization, i.e., the diff logs, are not sufficiently informative to guide the retrieval.

\subsection{Reproducible Build Related Work}
To the best of our knowledge, there have not been studies on localizing files that cause unreproducible builds. However, there have been studies that address the importance of reproducible builds. For example, Wheeler \cite{wheeler2005countering} describes a practical technique named diverse double compiling. By compiling the source files twice with different compilers, and verifying the compiled binaries, certain types of malicious attacks can be detected and prevented. According to Debian's documentation, this work partially motivates the reproducible builds practice \cite{debianabout}. Holler et al. \cite{holler2015evaluation} investigate the diverse compilation under embedded system, and experimentally quantify the efficiency of diverse compiling for software fault tolerance. Carnavalet and Mannan \cite{deCarnedeCarnavalet} conduct an empirical study, focusing on the reproducible builds in the context of security-critical software. Based on the experiments on the encryption tool TrueCrypt, they summarize the challenges of reproducibility in practice.
Ruiz et al. \cite{Ruiz:2015:RSA:2723872.2723883} address the reproducibility in cloud computing. They adopt the term reconstructable software, and propose a prototype to simplify the creation of reliable distributed software.

In this study, we focus on the localization task for unreproducible builds, which has not been addressed in the existing studies.

\section{Conclusions}
\label{conclusion}
In this study, we investigate the localization task for unreproducible builds. 
We present components that consider heuristic knowledge, similarity based information, as well as their integration as RepLoc.  For empirical validation, we create four categories of publicly available datasets with 671 unreproducible packages from Debian. Extensive experiments reveal that RepLoc is able to effectively localize the files that lead to unreproducible builds.  Furthermore, with the help of RepLoc, we successfully identified and fixed 6 new unreproducible packages from Debian and Guix.

For the future work, we are interested in the localization of problematic files for the tool-chain related issues. Also, 
inspired by the record-and-play techniques \cite{O'Callahan:2017:ERR:3154690.3154727} from the crash reproduction based debugging research \cite{Honarmand,xuancrash}, it would be interesting to leverage these techniques to detect more accurate correspondence between the build commands executed and the built binaries.

\section*{Acknowledgements}
This work is supported in part by the National Natural Science Foundation of China under Grants 61772107, 61722202, 61502345, and 61403057, and in part by the Fundamental Research Funds for the Central Universities under Grant DUT16RC(4)62.

\bibliographystyle{ACM-Reference-Format}
\bibliography{sigproc}  

\end{document}